# Are advanced methods necessary to improve infant fNIRS data analysis? An assessment of baseline-corrected averaging, general linear model (GLM) and multivariate pattern analysis (MVPA) based approaches

Maria Laura Filippetti[1], Javier Andreu-Perez[2], Carina de Klerk[1], Chloe Richmond[1], Silvia Rigato[1]

[1]Centre for Brain Science, Department of Psychology, University of Essex, Colchester, UK

[2]Centre for Computational Intelligence, School of Computer Science and Electronic Engineering (CSEE), University of Essex, Colchester, UK

## Abstract

In the last decade, fNIRS has provided a non-invasive method to investigate neural activation in developmental populations. Despite its increasing use in developmental cognitive neuroscience, there is little consistency or consensus on how to pre-process and analyse infant fNIRS data. With this registered report, we investigated the feasibility of applying more advanced statistical analyses to infant fNIRS data and compared the most commonly used baseline-corrected averaging, General Linear Model (GLM)-based univariate, and Multivariate Pattern Analysis (MVPA) approaches, to show how the conclusions one would draw based on these different analysis approaches converge or differ. The different analysis methods were tested using a face inversion paradigm where changes in brain activation in response to upright and inverted face stimuli were measured in thirty 4-to-6-month-old infants. By including more standard approaches together with recent machine learning techniques, we aim to inform the fNIRS community on alternative ways to analyse infant fNIRS datasets.

**Introduction**

The last 50 years have seen major advances in the development of methods and techniques that allow researchers to investigate the cognitive abilities of human infants, such as electroencephalography (EEG) and event-related potential (ERP), functional magnetic resonance imaging (fMRI) and more recently, functional near-infrared spectroscopy (fNIRS). This opportunity to investigate the functional brain development of infants in the first year of life, has led to significant progress in our understanding of infants' perceptual and cognitive abilities. One of the most promising new methods is fNIRS. This technique uses near-infrared light to detect changes in oxygenated (oxyHb; $HbO_2$) and deoxygenated (deoxyHb; HHb) haemoglobin concentrations (Lloyd-Fox, Blasi, & Elwell, 2010), which are a consequence of the increase in oxygen demand and are taken to indirectly reflect functional brain activation. Changes in concentration in $HbO_2$ and HHb during functional brain activation can be derived through light absorption measurements.

As with any other neuroimaging method, fNIRS is characterized by advantages as well as drawbacks. Over the last 20 years, its widespread use in infancy research can be attributed to the fact that this baby-friendly method is relatively cheap and easy to use, is less susceptible to movement artefacts compared to EEG, and allows for the spatial investigation of cerebral cortex activation in response to a variety of sensory stimuli (for reviews see Gervain et al., 2011; Issard & Gervain, 2018; Lloyd-Fox, Blasi, & Elwell, 2010), thus providing valuable insight into infants' neurodevelopment.

While the first uses of NIRS-based technology mainly involved clinical applications such as monitoring cerebral oxygenation in neonates (NIRS oximetry) (e.g. J. E. Brazy, Lewis, Mitnick, & Jöbsis vander Vliet, 1985; Jane E. Brazy & Lewis, 1986; Bucher, Edwards, Lipp, & Duc, 1993; Elwell et al., 2005; Wyatt, Delpy, Cope, Wray, & Reynolds, 1986), later, researchers started applying fNIRS to investigate functional brain activation in infancy (Meek et al., 1998; for a review see Gervain et al., 2011; Lloyd-Fox et al., 2010). This more recent research focused on the investigation of complex cognitive processes, such as speech processing and language development (e.g. Gervain, 2014; Gervain, Macagno, Cogoi, Peña, & Mehler, 2008; Issard & Gervain, 2017; S. Lloyd-Fox, Széplaki-Köllőd, Yin, & Csibra, 2015), the social brain network (e.g. S. Lloyd-Fox et al., 2015) and face processing (see Lloyd-Fox, Blasi, & Elwell, 2010 for a review of studies; Otsuka et al., 2007), and multisensory cues (Emberson, Richards, & Aslin, 2015; Filippetti, Lloyd-Fox, Longo, Farroni, & Johnson, 2015; Kersey & Emberson, 2017).

Despite its increasing use in developmental cognitive neuroscience, there is still little consensus on how to pre-process (Pinti, Scholkmann, Hamilton, Burgess, & Tachtsidis, 2019) and analyse fNIRS data. Indeed, the variety of NIRS systems, as well as the heterogeneity of statistical software and in-built data analyses scripts across infant laboratories, make it difficult if not impossible to establish the presence of consistent and reliable findings across experiments (although see Lloyd-Fox et al., 2017 for replication of social brain activation across different laboratories, populations, and set-ups; Benavides-Varela & Gervain, 2017; for replication of word learning; May, Gervain, Carreiras, & Werker, 2018 for replication of Gervain et al., 2008's rule learning paradigm). To this aim, the main purpose of the current study is to present the results of three fNIRS data analysis approaches: baseline-corrected averaging approach, GLM-based univariate analyses, and MVPA analyses, and discuss how the conclusions one would draw based on these different analysis approaches converge or differ.

Traditionally, fNIRS data has been analysed with baseline-corrected averaging techniques, which involve averaging together the neural response of blocks belonging to a given condition and performing a baseline-correction. The processed fNIRS data is then analysed using ANOVAs and/or paired sample channel-by-channel t-tests to compare experimental conditions ( Lloyd-Fox, Blasi, & Elwell, 2010). In this scenario, the time course data is first segmented into different time windows (i.e., baseline and conditions): the mean $HbO_2$ and HHb concentrations during the pre-experimental window (baseline) are then subtracted from the experimental trial window, to compute the mean haemodynamic concentration changes. The derived signals are then averaged across trials for each channel and condition and repeated measure analyses are conducted. Post hoc comparisons can then be made to control for false-positive activation using different multiple comparison methods. For example, fNIRS studies have used the False Discovery Rate (FDR) (Benjamini & Hochberg, 1995; Singh & Dan, 2006), Bonferroni correction (Dunn, 1961), spatially contiguous activation (Sarah Lloyd-Fox, Blasi, Everdell, Elwell, & Johnson, 2011; Southgate, Begus, Lloyd-Fox, di Gangi, & Hamilton, 2014), and Monte Carlo simulation (de Klerk, Bulgarelli, Hamilton, & Southgate, 2019). While the baseline-corrected averaging method is relatively easy to implement with fNIRS infant data, and avoids assumptions about the shape of the haemodynamic signal and its time course, it also disregards important timing information (Tak & Ye, 2014). Conversely, the General Linear Model (GLM) considers the whole-time course, thus proving a more powerful approach to the analysis of fNIRS data by making use of its good temporal resolution (Pinti et al., 2017). However, it is important to acknowledge that this timing information can partially be lost in the context of noisy infant data, where large segments of data sometimes have to be discarded due to infants' inattention or fussiness. While the GLM approach was originally implemented with fMRI data (Friston et al., 1996) it has been adapted for use with optical data (e.g., the NIRS-SPM toolbox; Ye, Tak, Jang, Jung, & Jang, 2009), based on the similarities between fMRI and NIRS designs and their reliance on the haemodynamic response. The GLM approach consists of modelling pre-specified regressors, which are then convolved with the expected haemodynamic response function (HRF) and fitted to the data (Pinti et al., 2017; Tak & Ye, 2014). However, the disadvantage of the GLM is that it assumes a predefined HRF, which e.g., can differ both within- and between- subjects and is not well established in newborns and young infants.

More recently, sophisticated techniques developed for fMRI data, such as multivariate methods like the multivoxel/multivariate pattern analysis (MVPA), have also been used to analyse fNIRS data. MVPA allows for the analysis of brain activity patterns that are distributed across cortical regions (Haxby, Connolly, & Guntupalli, 2014), by considering the relationship between multiple variables. For example, MVPA techniques allow researchers to investigate whether a pattern of fNIRS activation can discriminate between two or more conditions (Emberson, Zinszer, Raizada, & Aslin, 2017). In their recent work Emberson and colleagues (2017) applied MVPA to infant fNIRS data and successfully decoded channels that were responsive to their conditions of interest using a correlation-based decoding method to the recorded hemodynamic signals from an event-related design. Compared to standard univariate tests, MVPA has the potential to provide substantial details on aspects of the hemodynamic signal that could otherwise be missed using classic univariate tests (Emberson et al., 2017). First, the MVPA method can detect patterns of neural activation that are unique to the observation of one condition compared to another, hereby providing a higher level of analytic sophistication compared to standard univariate analyses. Indeed, compared to univariate tests which provide information about significant differences in activation between two conditions, MVPA can decode pattern of relatively subtle brain activity (for a discussion see Todd, Nystrom, & Cohen, 2013; Emberson et al., 2017). For example, both fMRI (Kok, Failing, & de Lange, 2014; Raizada & Kriegeskorte, 2010) and fNIRS (Emberson et al., 2017) studies have found multivariate methods to succeed in finding significant differences in distributed activation patterns and channels' contributions where univariate tests could not. Second, this method can reduce the effect of noise in observations by assimilating multiple noisy signals (hence resulting in a more powerful analysis than univariate tests) (Davis et al., 2014). Nevertheless, a problematic aspect of MVPA is that this type of approach does not account for certain types of confounds, which are appropriately addressed using more standard analyses (e.g. GLM; Todd et al., 2013). For example, because MVPA detects a brain signal that is individual-specific, the unique pattern of neural activity decoded in one condition might be due to a variety of individual confounds (e.g. individual differences on task performance) – rather than the condition itself. As a consequence, discrimination success from MVPA analysis could lead to false positives (Todd et al., 2013). However, some of the limitations associated with MVPA can be overcome by following certain procedures, such as avoiding overlapping of training and testing sets with data from the same run to reduce false positives (Mumford, Davis, & Poldrack, 2014). Overall, and considering the widespread use of a variety of analyses methods among the fNIRS community, comparing different approaches to analyse the same fNIRS dataset could give insight into the pitfalls and advantages of each method for exploring brain responses in infants.

**Our study**

With this registered report, we aimed to make an initial step towards establishing more robust fNIRS methods and data analysis, by comparing the results of three fNIRS data analysis approaches.

*Aim: Advance the statistical analysis of infant fNIRS data by illustrating and discussing the results of three fNIRS data analysis approaches: baseline-corrected averaging, GLM-based univariate analyses, and MVPA analyses. We investigated these different analysis methods in the context of a paradigm involving upright and inverted faces (i.e., face inversion effect).*

Specifically, thirty 4-to-6-month-old infants were presented with upright and inverted face stimuli in a block design. We relied on previous procedures used with infant fNIRS data to perform our baseline-corrected averaging (de Klerk, Hamilton, & Southgate, 2018) and GLM-based analyses (de Klerk et al, 2019; Pinti et al., 2017; Tak & Ye, 2014). For the multivariate analysis we used the MVPA method combined with machine learning techniques (by making use of support vector machine; SVM – a supervised learning model) which are frequently employed in multivariate analysis of adult fNIRS (Ichikawa et al., 2014) and fMRI data (Norman, Polyn, Detre, & Haxby, 2006). Within the fNIRS community, so far machine learning techniques have not been the standard data analysis approach. Therefore, this study implements different methods of analysis on the same infant fNIRS data, and discuss how results across these approaches might converge or differ.

The different analysis methods described above were tested using a face inversion paradigm. As shown in the adult literature (e.g., Minnebusch, Keune, Suchan, & Daum, 2010; Minnebusch, Suchan, Köster, & Daum, 2009; Tanaka & Farah, 1993), stimulus orientation effects measure configural structural representations of certain classes of stimuli. Such representations are thought to be the hallmark of functional specialisation for social stimuli such as faces. Previous infant EEG studies have found clear face-inversion effects (e.g. larger N290 in response to inverted faces compared to upright faces) suggesting that faces are processed configurally from birth and throughout the first year of life (e.g., Buiatti et al., 2019; de Haan, Pascalis, & Johnson, 2002; Halit, de Haan, & Johnson, 2003). Just a handful studies however, have examined the haemodynamic response to the inversion effect in infants (Ichikawa, Kanazawa, Yamaguchi, & Kakigi, 2010; Kobayashi et al., 2012; Otsuka et al., 2007), thus leaving the neural basis of face processing in young infants a critical point of investigation. Specifically, Otsuka et al. (2007) examined the classic face inversion effect by presenting static upright and inverted faces. Face stimuli were presented on alternating trials and images of vegetables were presented during the inter-trial intervals and used as baseline. Five- to eight-month-old infants showed some effect of face inversion; in the right lateral area of the infant brain the concentration of $HbO_2$ increased in response to upright faces (but not to inverted faces) compared to the vegetable baseline stimuli.

More generally, social stimuli such as faces or voices have been found to elicit canonical fNIRS responses earlier in development than non-social stimuli (e.g., Emberson et al., 2017; Grossmann, 2008; Lloyd-Fox et al., 2009; but see also Kobayashi et al., 2011), suggesting that these stimuli can be reliably contrasted to investigate the suitability of more sensitive and powerful analysis techniques for the use with infant fNIRS data. As the fNIRS infant literature on the face inversion effect is scarce, using the face inversion contrast in our study can provide further insight into the neural basis of face processing in young infants.

Within the temporal lobes, the superior temporal sulcus (STS) has been implicated in social perception in many previous studies (Allison, Puce, & McCarthy, 2000). Importantly, the perception of faces as well as eye gaze in static facial images has been shown to activate the STS region (Hoffman & Haxby, 2000; Itier & Taylor, 2004; Passarotti, Smith, DeLano, & Huang, 2007). Indeed, source localisation studies point towards the STS as one of the sources of the N170 component, showing larger intensities for faces compared to objects (Itier & Taylor, 2004). In our study, we minimise the inherent complexity of social stimuli (i.e. by using simple static human faces instead of complex dynamic social scenes), to target our STS area of interest.

It is not within the scope of this study to statistically assess similarities and differences among the three different analysis approaches here. Nevertheless, this investigation illustrates and discusses how results across different fNIRS analysis methods converge or differ. Given the large body of evidence suggesting that infants aged 6 months and under show enhanced activation to static social, face-like stimuli (Carlsson, Lagercrantz, Olson, Printz, & Bartocci, 2008; de Haan et al., 2002; Nakato et al., 2009; Otsuka et al., 2007), our study focuses on this age range (4-6 months) and uses simple visual stimuli, thus the outcome is specific to this protocol. The protocol associated with this Registered Report received in-principle acceptance (IPA) on 19 July 2019, prior to data collection and analysis. The approved Stage 1 manuscript, unchanged from the point of IPA, may be downloaded from https://osf.io/9u7xg.

**Predictions**

Based on the existing literature (Grossmann, Parise, & Friederici, 2010; Issard & Gervain, 2018; S. Lloyd-Fox et al., 2018), we expected the infants in our study to show a larger canonical hemodynamic response to upright human faces compared to inverted human faces. In particular, we expected that the STS area would show a significant difference in activation in response to upright, compared to inverted faces. Because changes in HHb signals in infants are generally smaller than changes in $HbO_2$ and do not always show the expected decrease over the course of a trial, a significant increase in $HbO_2$ in the absence of a significant decrease in HHb would be considered a positive result. Similarly, we would consider a positive result a significant decrease in HHb even if not accompanied by a concurrent significant increase in $HbO_2$ (cf. Lloyd-Fox et al., 2010).

Specifically, our hypotheses for the different analysis approaches are:

*Quality check (positive controls) for all our analyses (i.e. baseline-corrected averaging, GLM and MVPA)*:

**Sub-$H_0$:** Absence of significant differences in the hemodynamic response between inverted faces and images of vegetables (baseline condition) over the bilateral temporal arrays – and in particular over the STS. Our results would show no significant increase in $HbO_2$ and no significant decrease in HHb in response to inverted faces compared to baseline.

**Sub-$H_1$:** Significant differences in the hemodynamic response between upright face stimuli and images of vegetables (baseline condition) over the bilateral temporal arrays – and in particular over channels overlying STS areas. Our results would show a significant increase in $HbO_2$ and/or significant decrease in HHb in response to upright faces compared to baseline.

*Main hypotheses to test:*

**$H_{0UNIVARIATE}$:** In both baseline-corrected averaging and GLM analyses, channels over STS would not show a significant difference in activation in response to upright compared to inverted faces. This means that we expected no significant differences in $HbO_2$ or in HHb in response to upright compared to inverted faces over the bilateral temporal arrays.

**$H_{1UNIVARIATE}$:** In both baseline-corrected averaging and GLM analyses, channels over STS would show significantly greater activation in response to upright compared to inverted faces. This means that we expected a significantly greater increase in $HbO_2$ and/or a significantly greater decrease in HHb in response to upright compared to inverted faces over the bilateral temporal arrays.

**$H_{0MVPA}$:** Channels over STS would not differentiate between upright and inverted faces. This means that, we expected non-significant discriminatory patterns in $HbO_2$ and HHb in response to upright compared to inverted faces over the bilateral temporal arrays.

**$H_{1MVPA}$:** Channels over STS would differentiate between upright and inverted faces. This means that we expected significant canonical discriminatory patterns of activation (increase in $HbO_2$ and/or decrease in HHb) in response to upright compared to inverted faces over the bilateral temporal arrays.

While a right hemisphere dominance for processing of faces has been reported in previous fNIRS infant studies (Osaka et al., 2007), considering their very small sample sizes (ten babies) and the absence of clear lateralization effect for faces in infant EEG studies (de Haan et al., 2002; Halit et al., 2003), we did not have specific hypotheses with regard to hemispheric differences in face processing in our study.

**Participants**

Based on power analysis, we calculated that our sample size would need to comprise of 26 infants (please see details of power analyses in the *Sample Size Justification* section). Thirty-nine full-term, healthy four-to-six-month-old infants were recruited to participate in the study. According to the data exclusion criteria (please see Data Exclusion section), eight participants were excluded due to failure to look at the minimum 3 trials per experimental condition and one participant was excluded based on the NIRS data pre-processing (more than 30% of channels rejected by *enPruneChannels* function). Thus, the final sample was composed of 30 participants (13 female; $M_{age} = 162.03$ days, SD = 21.40 days).

Infants were recruited through the database of interested participants from the Essex Babylab and were born no more than a month before their due date, had no birth complications or major health problems, and no known hearing or vision difficulties. Caregivers were compensated with a £5 voucher for their visit and were given a token gift (e.g. a Babylab bodysuit/t-shirt, bib or tote bag).

*Sample Size Justification*

Justification of the sample size focused on demonstrating that the proposed Multivariate Pattern Analysis (MVPA; please see Data Analysis section) was able to provide a satisfactory performance in terms of overall accuracy to classify the patterns corresponding to each condition. The validity of our proposed analysis approach was determined using an effective pattern classification of a related developmental cognitive neuroscience fNIRS dataset where MVPA has been previously applied using correlation-based decoding approach (Emberson et al., 2017). The reliability of the MVPA analysis in our approach depends on the performance of a supervised learning model (linear SVM) with inputs consisting of activations patterns. The linear SVM classifier seeks a set of weights that best classify the activation patterns corresponding to each condition by the largest separative margin between them. For this, we have made use of the sample data of the original study by Emberson and colleagues (Emberson et al., 2017). This study investigated 5-6-month-old infants' neural processing of audiovisual stimuli (e.g. human faces and dimmed fireworks). Based on their results, Emberson and colleagues found differential informativeness of individual channels, and demonstrated that it is possible to use MVPA for decoding neural patterns as measured by fNIRS in developmental populations (Emberson et al., 2017).

To predict the sample size required for satisfactory performance of the MVPA in the current study, we used the method suggested by Figueroa and colleagues (Figueroa, Zeng-Treitler, Kandula, & Ngo, 2012). In this work, the required sample size estimated for each performance value (accuracy) was based on inverse power law model of the classifier's accuracies trend from a small annotated training set. The results of this analysis using Emberson's data are reported in Figure 1.

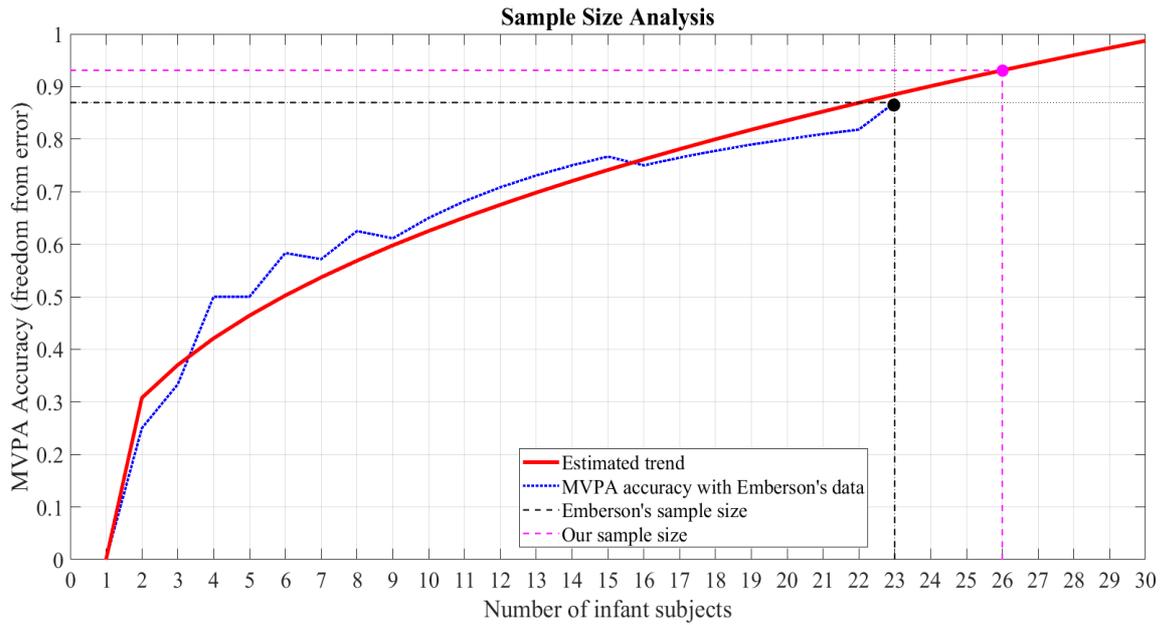

*Figure 1. Illustrative representation of the risk of error for the multivariate decoding of our MVPA approach using a regularised linear SVM-based method of Emberson's data (Emberson et al., 2017). The red solid line depicts an inverse power law curve with an increasing trend. The accuracy is defined as (True positives + True Negatives) / (True positives + True Negatives + False Positives + False Negatives) and for Emberson's study (dataset 1) with 23 subjects is 0.8696 (value displayed with a crossing black stripped line across both axes).*

From the analysis in Figure 4, we concluded that our MVPA based methodology is robust for the number of participants previously considered in related works and for our study (N = 26; see *Participants* section). The balance between the number of trials in Emberson's study is 50% for each condition. Our design also considered a 50-50% balanced number of trials for each condition for the sake of uniformity.

**Stimuli and Procedure**

The stimuli for the baseline period consisted of full-colour photo images of 5 vegetables, and those for the test period consisted of full-colour photo images of 5 female faces with a neutral facial expression, either in the upright or inverted position. The face stimuli were selected from the NimStim Face stimulus set (Tottenham et al., 2009); available at http://www.macbrain.org/resources.htm). To increase the likelihood of visual discrimination of stimuli (based on luminance, spatial frequency and colour), faces of multiple races and ethnicities were chosen (Emberson et al., 2017). The faces were cropped right below the neck. The LED monitor was 23.6" in size and located 90 cm from the participant's eyes.

The session began with a short, animated movie displaying animals, which we used to attract the infant's attention to the screen. As soon as the infant fixated on the monitor, trial presentation began. If necessary, occasional alerting sounds were played to draw the infant's attention back to the screen. To ensure that these sounds were balanced during the experimental session, each time the sound was used during the baseline trial, the following experimental trial also included a sound (Filippetti et al., 2015).

In each trial the five face images were shown in pseudo-random order at the rate of 1 Hz, to ensure that a given face image would not be presented more than twice within the same trial or in a row (Figure 2). Upright faces were displayed in half of the trials, and inverted faces in the other half of the trials. We used a pseudo-randomised order to ensure that a given condition was not presented more than three times in a row. Stimulus presentation continued until the infant became fussy or bored or until 30 trials had been presented as assessed by an experimenter who was monitoring their behaviour. The duration of the trials was fixed for 8 s. During the baseline trials, 5 images of vegetables were presented twice in random order at a rate of 1 Hz, for 10 s.

Infants were tested in a dimly lit and sound attenuated room, and were sitting on their parent's lap. Infants were encouraged to watch the stimuli displayed on the monitor. Parents were asked to refrain from talking and interacting with the infant during the stimuli presentation unless the infant became fussy. The computer played the stimuli through E-prime, and an HP laptop computer recorded the NIRS signal. The NIRS machine used pulsated LED emitters (NIRScout system, NIRx, BrainProducts). The whole testing session lasted about 10 minutes.

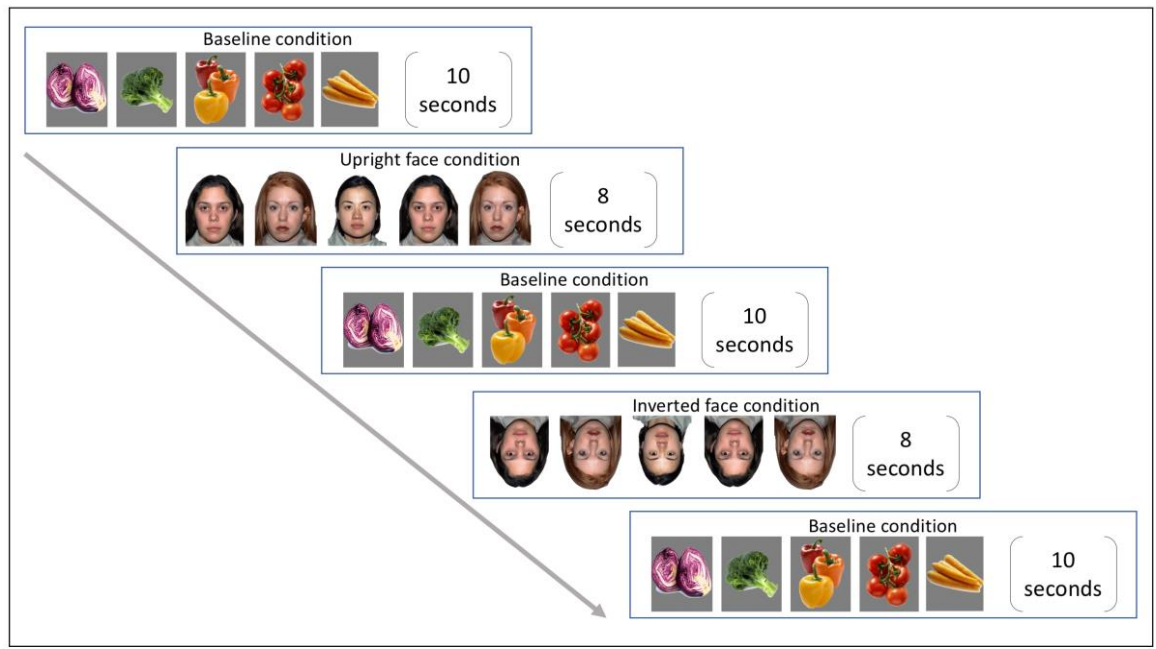

*Figure 2. Illustrative example of the block design adopted in our study. The experimental visual stimuli in both upright and inverted face conditions were presented in pseudo-random order for 8 seconds at a rate of 1Hz. In between trials, baseline stimuli were presented for 10 seconds and each of the 5 images were randomly displayed at a rate of 1Hz. Please note that this figure is for illustrative purpose only.*

**Data exclusion**

The decision to exclude a participating infant in subsequent analyses was made based on the following criteria:

- Refusing to wear the NIRS cap (none excluded in the study);

- Showing signs of distress (heavily fussing or crying) (none excluded in the study);

- Parental interference (e.g. talking and interacting with the infant during the stimuli presentation) that led to fewer than the required number of valid trials (none excluded in the study);

- Signal quality problems (due to pulling on the NIRS headgear, excessive movement or poor contact with the scalp; objective criterion: 30% of channels excluded) (one participant excluded based on this criterion);

- Experimental error, such as failing to record the fNIRS data or video, or misplacement of the source-detector montage arrangement (none excluded in the study);

- Looking time (eight participants excluded based on this criterion): Infants' looking behaviour was coded offline to ensure that trials were only included in the analysis if infants had watched at least 60% of the time in each trial as well as 30% of the pre- and post-stimuli baseline (Braukmann et al., 2018; S. Lloyd-Fox et al., 2013). The coder was blind to the study hypotheses.

**fNIRS recording and pre-processing**

fNIRS data was recorded using the NIRx NIRScout machine (source-detector separation: 2.5 cm; two wavelengths of 760 nm and 850 nm; sampling rate: approximately 10 Hz). The optical probe cap was placed on infants' heads targeting the temporal lobe areas of both hemispheres (left; LH, right; RH). Previous research using co-registration of fNIRS and MRI using a similar source-detector montage has demonstrated that it permits measurement of brain responses in cortical regions corresponding to IFG, STS, and TPJ areas (Lloyd-Fox et al., 2014).

The optical sensors were inserted into a stretchy EEG cap (Quick-cap, Compumedics Neuroscan) and were placed bilaterally on the infants' head using surface anatomical landmarks (inion, nasion, vertex and the bilateral preauricular points; see Figure 3 for approximate source-detector locations).

We used two different cap sizes (42 cm and 44 cm) depending on the participant's head circumference. We used a source-detector montage in line with previous research using co-registration of fNIRS and MRI (Lloyd-Fox et al., 2014). That is, for both caps we used two lateral arrays to form a total of 20 channels (Figure 3). We identified the location of responses from two measurements: 1) identification through external landmarks – photographs of the infant's head while wearing the headgear and 2) measurements of the infant's head circumference and distance between glabella, ears, vertex, pre-auricular points, and inion (Lloyd-Fox, Blasi and Elwell, 2010). In our sample, the average head circumference was 43.28 cm (SD = 1.29), and the average distance from the glabella to the ear above the pre-auricular point (T3/T4) was 11.35 cm (SD = 0.68 cm). The position of the channels over T3/T4 varied no more than 1cm along the axial plane across infants. We determined the approximate location of cortical regions based on previous co-registration research using a similar array with infants around the same age (Lloyd-Fox et al., 2014).

Data were pre-processed using Homer2, a Matlab software package (MGH-Martinos Center for Biomedical Imaging, Boston, MA, USA; Huppert, Diamond, Franceschini, & Boas, 2009). Data from the experimental session is automatically saved into ".nirs" format, compatible with Homer2. We followed a processing stream recommended for noisy data with few trials (typical of infant fNIRS data) (Brigadoi et al., 2014). Light intensity data was first converted to optical density and channels with raw intensities smaller than 0.001 V or bigger than 10 V were rejected (*enPruneChannels* function). Additionally, we checked that all channels with more than 20% of 'NaN' values (automatically selected segments of saturated data) were excluded from analyses and that all periods of 'NaN' values were excluded from the session. We manually excluded trials in which the infants were not attending and/or the caregivers were influencing the infants' looking behaviour (e.g. by talking– see exclusion criteria). Following the recommendations made by Brigadoi et al. (2014) motion artefacts were corrected using wavelet analyses with 0.5 times the interquartile range. To attenuate slow drifts and high frequency noise, we then band-pass filtered the data (high-pass: 0.01 Hz, low-pass: 0.80 Hz) using the default filtering options provided in HomER (Huppert et al., 2009). Infants for whom more than 30% of the channels were excluded due to weak or noisy signal were excluded from analysis. Finally, the data were converted to relative concentrations of oxygenated ($HbO_2$) and deoxygenated haemoglobin (HHb) using the modified Beer-Lambert law (Duncan et al., 1996) with a pathlength factor of 5.1 (Delpy et al., 1988). As in previous infant fNIRS studies, a total minimum number of 6 trials (3 trials per condition) was required to carry out the baseline-corrected averaging and GLM analyses (e.g.Lloyd-Fox et al., 2009, 2014; Southgate et al., 2014). For the MVPA approach, the total minimum number of trials for each subject was set to 16 (8 trials per condition) (Hua et al., 2004).[1]

---

[1] We originally planned to include 26 participants with at least 8 trials per condition. Unfortunately, the COVID-19 pandemic greatly disrupted the study and prevented further data collection. Therefore, we ran the MVPA both with our full sample of 30 participants who had a minimum 3 good trials per condition (N = 22 had at least 6 trials per condition), and with the 13 participants that had at least 8 good trials per condition. This change in data analysis was requested after data collection and pre-processing but before any further analysis or hypothesis-testing. The change received editorial approval on 5 May 2022.

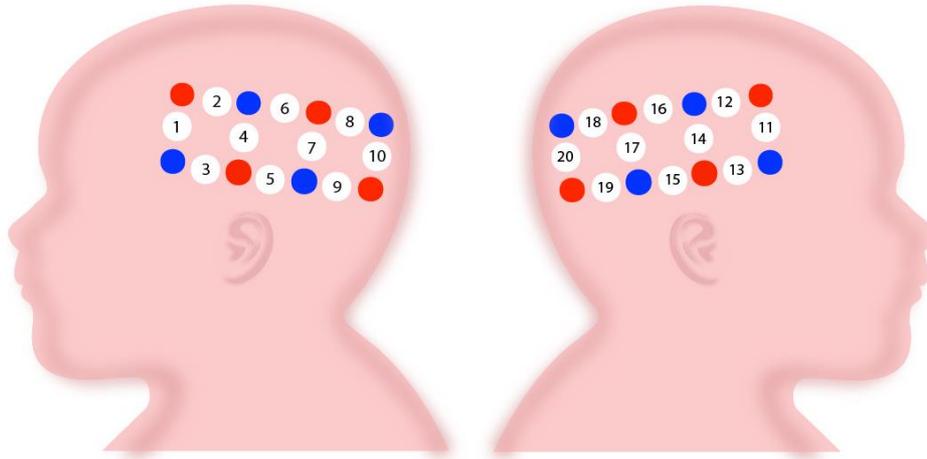

*Figure 3. Illustrative example of the source-detector configuration for the current study. The red dots represent the approximate source locations, whereas the blue dots represent the approximate detector locations. The white circles represent channels.*

**Data analysis**

Even though we are mainly expecting differences over STS areas, the array we used is slightly different from previous coregistration work (e.g. Lloyd-Fox et al., 2014), which makes it difficult to know exactly which channels to include in potential ROI analyses. Therefore, in the present study we performed whole-array analyses.

*Baseline-corrected averaging approach*

We computed relative changes in $HbO_2$ and HHb for 18-second long epochs starting 3 seconds before the onset of each trial and ending 7 seconds after trial offset. The mean $HbO_2$ and HHb concentrations during the 3-s pre-experimental window (baseline) was subtracted from the concentrations in the 15-s analysis period. The signals were then averaged across trials for each channel and condition. Following Lloyd-Fox et al. (2015), we first quantified the mean haemodynamic concentration changes during five 3-s sub-epochs following trial onset. Then, we performed repeated measures analyses on these five sub-epochs with the two conditions (upright face vs. inverted face) as within-subjects factors to identify channels that showed a significant $HbO_2$ increase and/or a significant HHb decrease from baseline when both conditions were considered together (as indicated by a main effect of time). To assess whether there were differences in the haemodynamic response between the two conditions (upright face vs. inverted face), we conducted repeated-measures analyses on each of these pre-selected channels that showed a significant haemodynamic response. Multiple comparisons corrections were done by controlling for the false discovery rate using the Benjamini-Hochberg procedure (Benjamini & Hochberg, 1995). We reported both uncorrected and corrected results, if at least the former was significant at $p < 0.05$. If the uncorrected result was significant, but the corrected one was marginal, we reported exact p values for the marginal result.

*General Linear Model (GLM) analysis*

Data analysis was conducted using a combination of custom code and the SPM-NIRS toolbox (Ye et al., 2009). For each infant, we constructed a design matrix with three regressors. The first regressor modelled the upright face trials (8 seconds), the second regressor modeled the inverted face trials (8 seconds), and the third regressor modeled the baseline trials (10 seconds). We set excluded trial periods to zero, thus effectively removing them from the analyses. These regressors were convolved with the standard canonical haemodynamic response function without derivatives to make the design matrix (Friston, Ashburner, Kiebel, Nichols, & Penny, 2011), which we then fitted to the data using the general linear model as implemented in the SPM-NIRS toolbox (Ye et al., 2009). For each of the regressors and for each infant, we obtained beta parameters that were then used to calculate a contrast between the conditions of interest for each infant. To ensure statistical reliability of channel activation, we used FDR correction using the Benjamini-Hochberg procedure (Benjamini & Hochberg, 1995) as detailed in the baseline-corrected averaging approach.

*Multivariate Pattern analysis (MVPA)*

Data analysis was conducted using custom in-house scripts (code is available via GitHub). Multivariate decoding accuracy was estimated as basis of the prediction of the SVM that was trained with the following target labels for each vector: 1) upright vs. baseline; 2) inverted vs. baseline; 3) upright vs. inverted. The multichannel vector of 1st level GLM $\beta$ coefficients for each condition was used as features (i.e. using the output from the GLM with one regressor per condition and channel) in our subsequent MVPA analysis. In order to validate the generalization of the MVPA results, testing and training sets were generated in a subject-wise repeated 5-fold cross-validation procedure with non-overlapping subjects and outcome values were reported as the average across cross-validation test. The values of $\beta$ from the missing channels were estimated by applying the Bayesian Principal Component imputation method, using as reference the $\beta$ value of the non-missing channels within the same area (Audigier, Husson, & Josse, 2016). A support vector machine (SVM) was used as binary classifier as applied before on MVPA analysis and with fNIRS signals (Andreu-Perez, Leff, Shetty, Darzi, & Yang, 2016; Mourão-Miranda, Friston, & Brammer, 2007). Using SVM with a linear kernel is a recommended approach for MVPA as this simpler model is less prone to overfitting (i.e. modelling noise along with relevant information). Additionally, we also fitted the SVM with L2 (ridge) regularization that aims to keep the model weights small, hence reducing its complexity and avoiding overfitting. A search guided by a genetic algorithm (Xue et al. 2016; Yang and Honavar 1998) was used to find the subset of the most determinant channel betas that maximises the MVPA accuracy. Parametrization for the SVM was estimated from the cross-validation procedure within the training set only. We applied a dataset resampling procedure via randomization approach (Monte Carlo guided permutation) to perform group level analysis with a nested 5-fold cross validation (Etzel, 2015). The classification analysis was run multiple times with the permuted labels to obtain a null distribution or chance performance distribution, and significance of the observed results was assessed by the empirical p-value for the null distribution. The same procedure was performed for the sensitivities, i.e., scores of the classifier for each feature. A schematic representation of all the processing steps leading to the MVPA analysis from the raw light intensity (Volts) measures, is presented in Figure 4.

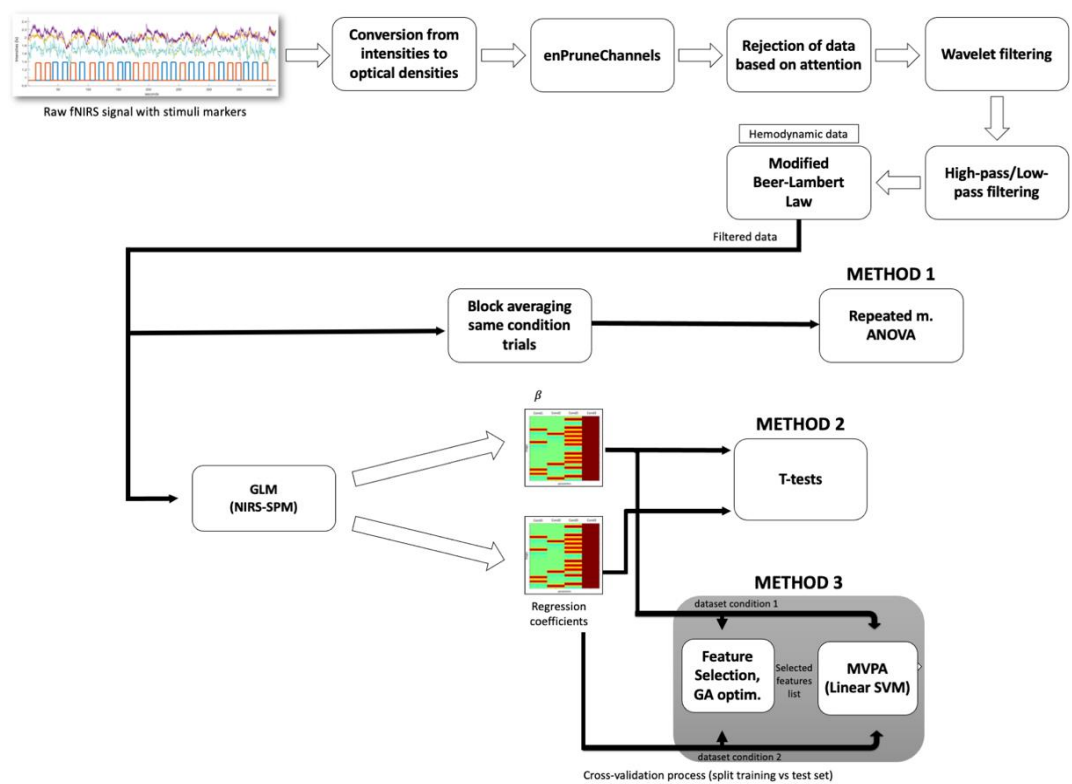

*Figure 4. Schema of the data processing stack including all steps towards the three analysis methods (baseline-corrected averaging: Method 1; GLM: Method 2; MVPA: Method 3). Intensity data were filtered and converted to concentration of each haemoglobin chromophore. The haemodynamic data were then transferred to a GLM process to estimate the beta*

*regression coefficients for each channel and stimulus condition. The regression coefficients corresponding to each condition were used as input features for feature selection via Recursive Feature Elimination with cross-validation (RFECV) and MVPA analysis, all encapsulated within a repeated k-fold cross-validation process to split between the training and test sets.*

**Results**

*Baseline-corrected averaging results*

The initial analyses identified three channels that showed a significant haemodynamic response during the trial period compared to the baseline period when both conditions were considered together (as evidenced by a significant main effect of time). These channels showed a significant haemodynamic response through an increase in $HbO_2$ between the baseline period and the trial period: channel 8, $F(1.976, 61.655) = 5.567$, $p = .0006$, channel 18, $F(1.671, 34.184) = 7.448$, $p = .003$, channel 20, $F(2.249, 52.135) = 3.092$, $p = .046$. Channel 8 also showed a significant haemodynamic response through a decrease in HHb between the baseline and trial, $F(2.285, 57.395) = 4.638$, $p = .010$. For channel 18, there was a significantly greater $HbO_2$ response to the Upright compared to the Inverted face condition (main effect of condition, $F(1,29) = 4.838$, $p = .036$; see Figure 5) indicating a greater $HbO_2$ response to the Upright face condition throughout the analysis period. However, this channel did not survive multiple comparisons correction, FDR-corrected $p = .108$. Please see Table 1 for a summary of preregistered results.

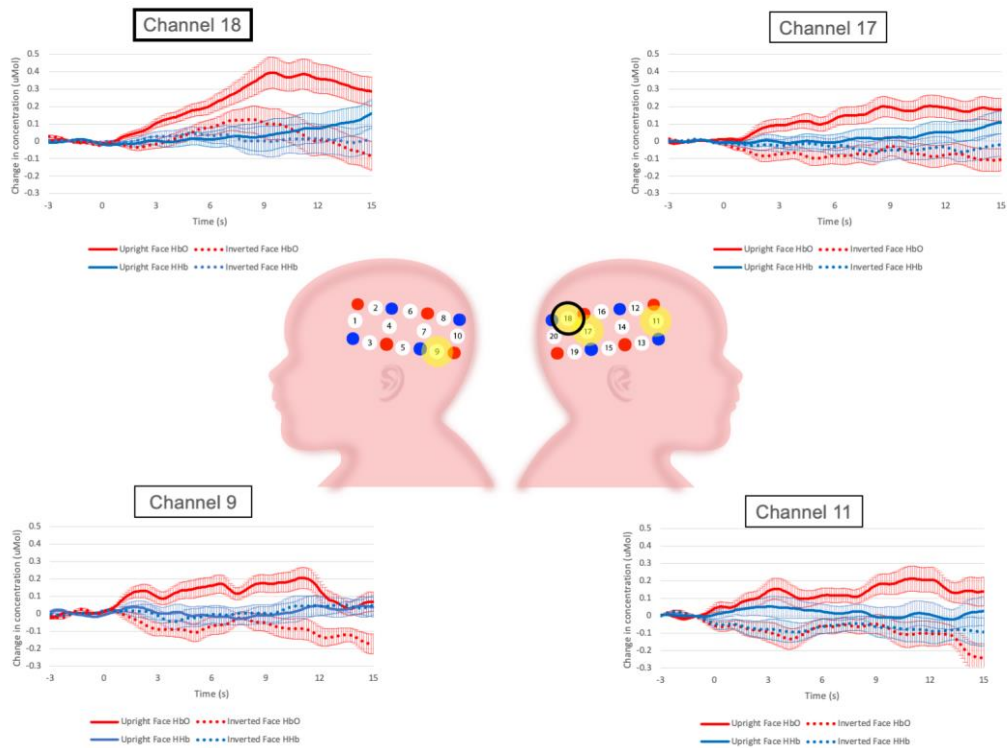

*Figure 5. Hemodynamic response to the upright (solid line) and inverted (dashed line) face conditions for channels 18 (top left), 17 (top right), 9 (bottom left), and 11 (bottom right). These channels are also highlighted in yellow in the array image. In the baseline-corrected analysis, channel 18 (in bold) was the only channel that displayed a main effect of time and a main effect of conditions. The significant responses are illustrated in red ($HbO_2$) and blue (HHb).*

## General Linear Model (GLM) results

*t* Tests on the betas for the contrasts between the individual conditions and baseline showed no significant increase in $HbO_2$ and/or significant decrease in HHb in response to upright faces compared to baseline or in response to inverted faces compared to baseline. Thus, our GLM output data did not pass the initial quality check. *t* Tests on the betas for the contrast between the upright and inverted condition revealed two channels that were sensitive to face orientation - one channel over the left temporal parietal cortex (channel 8) and one channel over the left inferior frontal cortex (channel 1). Channel 8 showed a significantly greater hemodynamic response (based on $HbO_2$) for the Upright compared to the Inverted face condition, t(28) = 2.268, *p* =.031. However, this channel did not survive multiple comparisons correction, FDR corrected $p = .56$. Channel 1 showed a significantly greater hemodynamic response (based on HHb) for the Upright compared to the Inverted face condition, $t(26)= 2.335$, $p = .028$. However, this channel did not survive multiple comparisons correction, FDR corrected $p= .62$. No channels showed a significantly greater response in the Inverted condition compared to the Upright face condition. Please see Table 1 for a summary of preregistered results.

*Multivariate Pattern analysis (MVPA) results*

We ran the MVPA with both the full sample of 30 participants with a minimum of 3 trials per conditions (dataset 1 thereafter), and with the 13 participants that had at least 8 trials per condition (dataset 2 thereafter). For dataset 1, decoding accuracy (50%) was not statistically significant for $HbO_2$ during the trial period compared to the baseline period; upright vs baseline: $p = .55$; inverted vs baseline: $p = .50$. Similarly, decoding accuracy (58%) was not statistically significant for $HbO_2$ between upright vs inverted conditions, $p = .13$. Decoding accuracy was also not statistically significant for HHb in all comparisons: upright vs baseline: accuracy = 55%, $p = .29$; inverted vs baseline: accuracy = 50%, $p = .50$; upright vs inverted: accuracy = 56.67%, $p = .23$. Thus, while these accuracy scores were numerically above 50%, they did not exceed chance level.

For the $HbO_2$ decoding of Upright vs Inverted condition, we also extracted channel decoding relevancy (Figure 6) as revealed by the absolute value of the coefficients of the linear SVM-based method used for the MVPA analysis. Channels 15 and 17 on the left hemisphere displayed larger coefficients than the rest of the channels on both hemispheres. However, these channels were not significantly more relevant than the others, thus suggesting that the decoding of 58% accuracy is achieved considering the whole multivariate set of channels. Please see Table 1 for a summary of preregistered results.

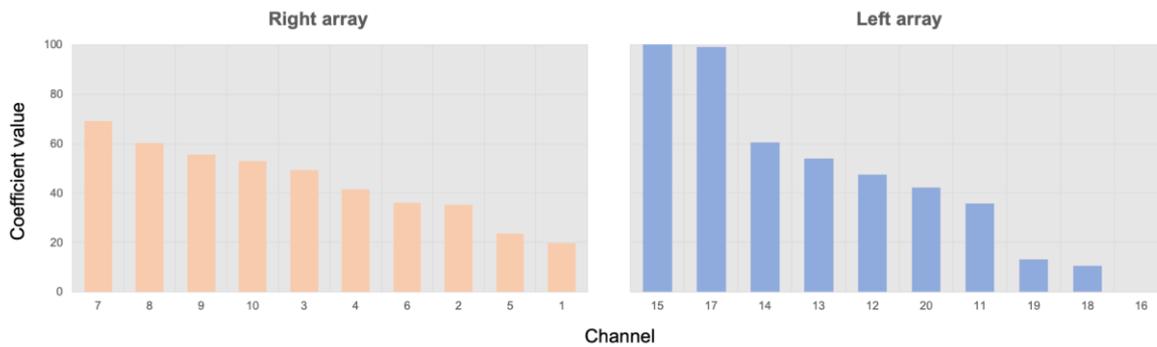

*Figure 6. Channel decoding relevancy revealed by the coefficients of the base model Linear SVM for MVPA with dataset 1. Each channel is grouped by side of side of the array.*

We found similar results for dataset 2. Decoding accuracy was not statistically significant for $HbO_2$ during the trial period compared to the baseline period; upright vs baseline: accuracy = 46%, $p = .65$; inverted vs baseline: accuracy = 61%, $p = .16$. The group model for $HbO_2$ in the comparison upright vs inverted yielded 61% accuracy. However, this accuracy score did not robustly exceed chance, $p = .16$. Decoding accuracy was also not statistically significant for HHb in all comparisons: upright vs baseline: accuracy = 58%, $p = .28$; inverted vs baseline: accuracy = 49%, $p = .41$; upright vs inverted: accuracy = 50%, $p = .45$. Overall, we therefore found that the model did not manage to establish an accurate decoding whether the infants in our sample were watching either an upright or inverted face stimulus.

For the $HbO_2$ decoding of Upright vs Inverted condition we also extracted channel decoding relevancy (Figure 7) as revealed by the absolute value of the coefficients of the linear SVM-based method used for the MVPA analysis. Channel 7 on the right hemisphere displayed a larger coefficient than the rest of the channels on both hemispheres. However, this channel was not significantly more relevant than the others, thus suggesting that the decoding of 61% accuracy is achieved considering the whole multivariate set of channels.

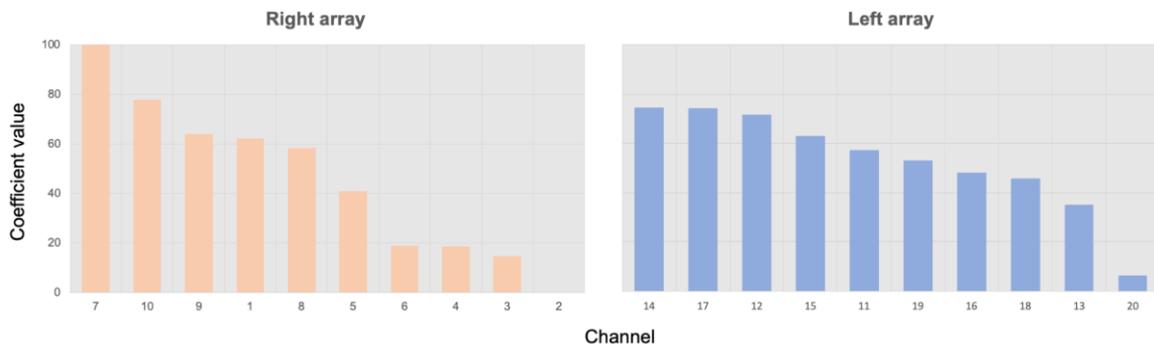

*Figure 7. Channel decoding relevancy revealed by the coefficients of the base model Linear SVM for MVPA with dataset 2. Each channel is grouped by side of the array.*

**Exploratory analyses**

*Baseline-corrected averaging results*

The requirement of the preregistered univariate analysis was for the channels to display a main effect of time, that is to show a significant haemodynamic response during the trial period compared to the baseline period when both conditions were considered together. We set this prerequisite to reduce the number of comparisons, however it is likely that some channels were sensitive to face orientation despite not showing evidence of a significant hemodynamic response when both conditions were considered together. Therefore, we also report the results on main effect of condition (as indexed by an increase in $HbO_2$ or a decrease in HHb).

Three additional channels overlying the posterior temporal areas bilaterally (channel 9 and channel 17) and the right anterior temporal area (channel 11) showed a significant effect of condition based on $HbO_2$, indicating a greater response to the Upright compared to the Inverted condition. However, these channels did not show a significant effect of time or an interaction between time and condition; channel 9: $F(1,29)= 4.317$, $p =.047$, FDR corrected $p =.235$; channel 11: $F(1,27) = 7.541$, $p =.011$, FDR corrected $p =.22$; channel 17: $F(1,28) =$

7.242, $p$ =.012, FDR corrected $p$ =.12). None of the channels showed a significant effect of condition based on HHb. For a summary of exploratory results, please see Table 2.

*General Linear Model (GLM) results*

The GLM-based univariate analysis did not find evidence for a significant hemodynamic response between the individual experimental conditions and baseline. Given that the HRF is not well established in young infants, it is possible that the estimation of beta coefficients from this approach lacked sensitivity. Therefore, we ran an exploratory GLM analysis using time and dispersion derivatives when modelling the HR, thus taking into account variability in the onset time and shape of the hemodynamic response that are common in infant data (see Table 2 for a summary of exploratory results). t Tests on the betas for the contrasts between the individual conditions and baseline showed a significant increase in $HbO_2$ in response to baseline compared to the upright faces in channel 12 (t(29)= -2.359, $p$ = .025, FDR corrected $p$ = .50). There was also a significant increase in $HbO_2$ in response to baseline compared to inverted faces in channels 12 (t(29) = -2.494, $p$ = .019, FDR corrected $p$ = .19) and 14 (t(28) = -2.735, $p$ = .011, FDR corrected $p$ = .22). Channel 5 showed a significant decrease in HHb for the Upright compared to the baseline condition, t(29) = 2.759, $p$ = .010, FDR corrected $p$ = .20), as well as for the Inverted compared to the baseline condition, t(29) = 2.162, $p$ = .039, FDR corrected $p$ = .39. Additionally, channel 1 showed a significant decrease in HHb in response to baseline compared to the inverted faces, t(29) = -3.281, $p$ = .001, FDR corrected $p$ = .06. Please note that none of the above effects survived FDR correction.

t Tests on the betas for the contrast between the upright and inverted condition revealed that channel 8 and channel 17 were sensitive to face orientation – although these effects did not survive FDR correction. Both channels showed a significantly greater hemodynamic response (based on $HbO_2$) for the Upright compared to the Inverted face condition, channel 8: t(28) = 2.105, p =.044, FDR corrected $p$ = .44; channel 17: t(28) = 2.293, $p$ =.030, FDR corrected $p$ = .60. No channel showed a significantly greater response in the Inverted condition compared to the Upright face condition. No channel showed a significant decrease in HHb in the contrasts between Upright and Inverted face condition.

*Multivariate Pattern analysis (MVPA) results*

Our GLM analysis did not pass the initial quality checks, showing no significant increase in $HbO_2$ and/or significant decrease in HHb in response to upright faces compared to baseline. This suggests that our GLM model and, consequently our beta-based MVPA, may have lacked sensitivity and may not have been optimal for the analysis of the current dataset. Therefore, we ran additional (non-preregistered) MVPAs using 1) baseline-corrected responses instead of beta coefficients and 2) beta coefficients derived from the GLM using time and dispersion derivatives (see Table 2 for a summary of exploratory results).

If our MVPA was affected by the estimation of betas, then we should observe that MVPA using baseline-corrected responses would provide results in line with our baseline-corrected univariate approach. The MVPA approach used was similar to the beta-based approach that we preregistered (please see *Multivariate Pattern analysis (MVPA)* section) with the exception that in this model our features were the multichannel vector of baseline-corrected averages for each condition over the time window 0 s - 12 s of each block. In addition, the MVPA was configured to select best multivariate combinations with a minimum of 2 channels. Data analysis was conducted using custom in-house scripts (code is available via GitHub).

We ran the MVPA with the full sample of 30 participants. The group model for $HbO_2$ in the comparison upright vs inverted yielded 72% accuracy. This accuracy score was statistically significant, $p = .009$, with the multivariate combination of channels 8 (decoding relevancy, $p = .07$) and 17 (decoding relevancy, $p = .009$) producing high decoding power, although only the decoding relevancy of channel 17 was significant. Decoding accuracy was also statistically significant for HHb, with a decoding accuracy of 75%, $p = .009$. The model showed that a larger set of channels contributed to the high decoding power, although only channel 11 relevancy was consistently significant ($p = .03$). Overall, these results show the expected significant canonical discriminatory patterns of activation (increase in HbO2 and/or decrease in HHb) in response to upright compared to inverted faces over the bilateral temporal arrays.

To examine whether decoding accuracy would benefit from betas derived from a GLM approach that uses time and dispersion derivatives, we ran an additional (non-preregistered) MVPA with dataset 1 (N = 30 participants). In this model, the set of best channel's beta-coefficients is selected by means of recursive feature elimination (REF). This method starts with a base model 'linear-SVM' including all channels' beta-coefficients ('feature') and recursively prunes features and checks for the improvement or deterioration of the decoding accuracy upon removal of the features (Guyon et al, 2002). The automatic selection of the length of the optimal feature set is achieved by wrapping REF with a cross-validation procedure, where at every step of removal a decoding score is computed based on cross-validation. The step that yields the higher score is considered the best number of features. A subsequent RFE is ran in order to select the best set of features of that number. With the based model trained with the best set of features, a permutation test is performed to estimate the empirical p-value of decoding accuracy ('score'), and feature importance ('coefficients').

Using this model, we found that decoding accuracy was statistically significant for $HbO_2$ during the trial period compared to the baseline period; upright vs baseline: accuracy = 70%, $p = < .01$; inverted vs baseline: accuracy = 67%, $p = < .01$. Similarly, decoding accuracy was statistically significant for HHb for the contrast upright vs baseline: accuracy = 66%, $p = .0198$ – but not for the contrast inverted vs baseline: accuracy = 58%, $p = .089$.

The group model for $HbO_2$ in the comparison upright vs inverted yielded 80% accuracy. The model showed that a large set of channels contributed to the high decoding power, with channels 5 ($p=0.0218$), 13 ($p = < .01$), 15 ($p = < .01$), 18 ($p = < .01$), 19 ($p = < .01$), and 20 ($p = < .01$) being consistently significant. Decoding accuracy was also statistically significant for HHb, with a decoding accuracy of 72%, $p = < .001$, with the multivariate combination of channels 16 (decoding relevancy, $p =< .01$) and 18 (decoding relevancy, $p = 1.941$) producing high decoding power, although only the decoding relevancy of channel 16 was significant. The results of this MVPA using beta coefficients from the GLM with derivatives show significant discriminatory patterns in response to upright compared to inverted faces over the right temporal array.

*Table 1* **Results of the preregistered analyses**. *Summary of data analyses results across the baseline-averaging, GLM, and MVPA methods.*

| | Preregistered analyses | | |
|---|---|---|---|
| | Baseline-averaging | GLM (Anova) | MVPA with beta (GLM) |
| **Pass quality check?** *(Sub-$H_0$/Sub-$H_1$)* | Sub-$H_1$ | Neither | Neither |
| **$H_1$** *(upright vs inverted)* | Yes | Yes | No |
| **Significant channels** *(upright vs inverted)* | Channel 18 ($HbO_2$) [not surviving multiple comparison correction] | Channel 8 ($HbO_2$) and channel 1 (HHb) [not surviving multiple comparison correction] | None |

*Table 2* **Results of the exploratory analyses**. *Summary of data analyses results for the baselin-averaging, GLM using time and dispersion as derivatives, baseline-corrected MVPA, and MVPA using betas derived from GLM with derivatives.*

| | Exploratory analyses | | | |
|---|---|---|---|---|
| | Baseline-averaging | GLM with derivatives | Baseline-corrected MVPA | MVPA with beta (GLM with derivatives) |
| **Pass quality check?** *(Sub-$H_0$/Sub-$H_1$)* | Sub-$H_1$ | Sub-$H_0$ | n/a | Yes |
| **$H_1$** *(upright vs inverted)* | Yes | Yes | Yes | Yes |
| **Significant channels** *(upright vs inverted)* | Channels 9, 11, and 17 ($HbO_2$) | Channels 8 and 17 ($HbO_2$) [not surviving multiple comparison correction] | Channel 17 ($HbO_2$) and channel 11 (HHb) | Channels 5, 13, 15, 18, 19, 20 ($HbO_2$), and channel 16 (HHb) |

## Discussion

Over the last few decades fNIRS research has been growing rapidly in the field of developmental cognitive neuroscience. Possibly partially as a result of the rapid developments in this field, there are many different approaches to the analysis of developmental fNIRS data. In the current registered report we aimed to compare and contrast three fNIRS data analysis approaches, namely baseline-corrected averaging, GLM-based univariate analyses, and MVPA analyses, in the context of a paradigm involving upright and inverted faces (i.e., the classic face inversion effect) with infants. We discuss the results of each approach separately before considering how these converge or differ.

The baseline-corrected averaging approach showed that only channel 18, approximately overlying the right posterior STS region (based on the standardised scalp surface fNIRS map from Lloyd-Fox et al., 2014), displayed a significantly greater $HbO_2$ response to the Upright compared to the Inverted face condition while also showing a significant main effect of time. While the FDR used in the present study is considered to be highly conservative for infant fNIRS analyses (Filippetti et al., 2015; Lloyd-Fox et al., 2017), it is important to note that channel 18 did not survive multiple comparisons correction and thus this result should be treated with caution. We also identified three channels situated over the posterior bilateral region of the array (channels 17 and 9) and one channel situated over the right anterior region (channel 11) that were sensitive to face orientation but that did not show evidence of a significant hemodynamic response (as indexed by an increase in $HbO_2$ or a decrease in HHb) when both conditions were considered together (a prerequisite we set to reduce the number of comparisons). We identified the location of these channels as lying over bilateral posterior STS region and the right inferior frontal gyrus (Lloyd-Fox et al., 2014). Overall, these results are in line with Otsuka et al. (2017) who found that five-to-eight-month-old infants' brains showed a significant increase in $HbO_2$ over the right temporal lobes in response to upright (but not inverted) faces.

The GLM-based univariate analyses revealed that one channel situated over the left posterior region of the array (channel 8; corresponding to the left posterior STS region) displayed a significantly greater $HbO_2$ response to the Upright compared to the Inverted face condition. This result corroborates previous research pointing to the STS as a key brain region implicated in social perception (Allison, Puce, & McCarthy, 2000). However, given that this effect did not survive correction for multiple comparisons, and that a significant difference over a single channel (in the absence of significant effects over adjacent channels) may lack statistical reliability (Lloyd-Fox, Blasi, Everdell, Elwell, & Johnson, 2011), this could be a false positive finding. Additionally, given that we did not find evidence for a significant hemodynamic response to the upright condition compared to baseline, we have reason to question the sensitivity of the GLM analyses. As outlined in the Introduction, the GLM methods makes assumptions about the shape and timing of the HRF (Tack & Ye, 2014). Given that the HRF is not well established in young infants, assumptions about the HRF are susceptible to model misspecification (Yücel et al., 2021). A potential solution is to use time and dispersion derivatives when modelling the HR, which takes into account variability in the onset time and shape of the hemodynamic response. Our exploratory GLM analysis using time and dispersion derivatives showed that channel 8 and channel 17, approximately overlying the posterior bilateral region of the array, were sensitive to face orientation – although these effects did not survive FDR correction. Overall and despite the potential lack of sensitivity of our GLM model, these results are in line with the baseline-corrected analysis and point towards channels lying over the STS region of the infant brain being sensitive to face orientation.

Unsurprisingly given these potential issues with the GLM approach, the MVPA analysis using beta coefficients was not able to distinguish between the two stimulus conditions in our fNIRS data with the anticipated level of accuracy (*Figure 1*). Higher decoding accuracy was achieved by the MVPA based on baseline-averaging and the MVPA based on beta coefficients derived from the GLM with derivatives that we ran as exploratory (non-preregistered) analyses. The MVPA based on baseline-averaging showed some convergent results with the baseline-averaging univariate approach, thus reinforcing the idea that our GLM model, and consequently our beta-based MVPA, may have lacked sensitivity and may not have been optimal for the analysis of the current dataset. The MVPA approach based on beta coefficients derived from the GLM with derivatives yielded 80% decoding accuracy for the face orientation contrast. This model showed that a large set of channels lying over the right STS region of the infant brain jointly contributed to the high decoding power. These results suggest that beta-based MVPA approaches should not rely on a priori estimation of beta coefficients unless the experimental design and/or population are conducive to eliciting a clear canonical HR. Other multivariate methods not dependent on beta-weight regression or HRF fitting may also be helpful (Emberson et al., 2017; Kiani et al., 2020).

The main aim of the present study was to analyse our fNIRS data using three different approaches and discuss how these results would converge or differ. The preregistered univariate analyses (baseline-averaging and GLM-based approaches) provided some convergent results. While no single channel showed a significantly greater response to the Upright compared to the Inverted face condition in both analysis approaches, these analyses did seem to point towards channels lying over the STS region of the infant brain being sensitive to face orientation. This finding suggests that there are univariate differences between upright and inverted face conditions.

Our preregistered multivariate analysis was not able to distinguish among conditions and thus these results diverge from the baseline-averaging and GLM-based univariate approaches. This is in contrast with Emberson and colleagues (2017)'s study which also used a MVPA approach and reported significant decoding of unimodal and bimodal (audiovisual) stimuli with smaller sample size and trial numbers (N = 25 and 6.9 trials in Dataset 1; N = 26 and 4.9 trials in Dataset 2; Emberson et al., 2017) than the one employed in the current study (N = 30, 8.2 trials). However, Emberson's study presented two important differences compared to our study: 1) it used a multivariate approach relying on correlations among trials rather than beta coefficients and 2) it involved a perceptual discrimination between distinct visual and auditory stimuli whereas the present study examined configural structural representations of faces, which could arguably be considered a more subtle contrast. Nonetheless, our MVPA exploratory analysis using baseline-corrected data (instead of betas) and using betas derived from GLM with derivatives, found significant decoding, thus converging with Emberson et al. (2017)'s findings that a MVPA approach to fNIRS infant data can be implemented even with more subtle stimulus contrasts. Another consideration in relation to Emberson et al. (2017)'s work is that their MVPA model focused on a subset of 10 channels whereas our model considered the whole fNIRS array of 20 channels. Indeed, in their study, the authors found greater decoding accuracy for subsets of small number of channels (2 to 10 channels). However, we think it is unlikely that focusing on a subset of channels in our MVPA would have significantly improved the model given that a feature selection pipeline was already implemented in the MVPA analysis and that the absolute values of the linear SVM coefficient intrinsically determine feature relevance (Guyon & Elisseeff, 2006). That is, the MVPA SVM-based method we employed automatically weighs the contribution of the channels for decoding so any low-weighted input would have had a minimal impact on the model.

From the observations made thus far it is possible to outline some considerations that developmental researchers may wish to take into account when designing, and planning data analysis of a fNIRS study. First, although the GLM approach may not have been suited for the analysis of the current dataset, it does provide clear benefits for fNIRS studies in which the researcher wants to control for additional regressors (e.g., other physiological measures, behavioural responses, etc.). A benefit of GLM approaches is that these variables can be added to the model to estimate their effect in the fNIRS signal – which is not possible when employing a baseline-corrected averaging approach. In addition, GLM may also be particularly suitable for data collected with slightly older developmental populations (i.e., children) when the HRF becomes more similar to the canonical response. With younger infants, GLM approaches that take into account HRF variability should be considered. Our GLM using time and dispersion derivatives proved promising, but there are other methods that can extract intricate hemodynamic functions from fNIRS (e.g., using a Finite Impulse Response Filter (FIR) approach (Huppert, 2016; Pinti, Scholkmann, Hamilton, Burgess, & Tachtsidis, 2019)). However, the high uncertainty in the shape of the HRF in developmental populations (Gemignani & Gervain 2021), vasoconstriction dynamics and non-stationarities of neurovascular coupling, render analysis standards sustained in waveform fitting an open question. This report also shows that assumption-free analysis methods can be useful for this field of research.

Second, and relatedly, more refined MVPA approaches might be particularly useful in modelling the uncertainty in the input data – especially in the context of infant data which is often characterized by inter-subject and/or intra-subject variabilities. For example, MVPA based on eXplainable Artificial Intelligence (XAI) can describe the activation level of cortical regions based on conceptual labels outlining multivariate contributions between brain regions for stimulus processing (Kiani et al., 2020). Such data-driven approach may be based on time-discretised baseline-corrected responses or any other signal descriptor - overcoming the need of determining *a priori* the shape of the HRF (Gemignani, 2018) - and may be useful for identifying patterns of intra- and inter-regional interactions (Andreu-Perez et al., 2021; Kiani, Andreu-Perez, Hagras, Filippetti, & Rigato, 2020; Kiani, Andreu-Perez, Hagras, Rigato, & Filippetti, 2022).

With this registered report, we contribute to the body of recent work aimed at improving the analysis of fNIRS data by building a consensus of best practices (e.g., (Gemignani & Gervain, 2021; Pinti et al., 2019; Yücel et al., 2021)). While it was not within the scope of the study to provide a definite solution to the design and analysis of infant fNIRS data, the added value of this work is that it provides the first systematic comparison of how different types of analysis approaches can affect the results. Importantly, as part of the registered report process, the significance of the research and robustness of the methods are evaluated - and approved - via peer-review before any data collection takes place. The report details our approach to determining the appropriate statistical power for this study and follows well-established pre-processing pipelines to ensure good data quality. Therefore, it is highly unlikely that our results can be explained by small sample size or inadequate data quality.

Overall, future studies should carefully consider the caveats outlined in the present work when carrying out fNIRS developmental research using univariate and/or multivariate approaches. In a recent comment published in Nature, Wagenmakers, Sarafoglou and Aczel (2021) suggested that performing multiple analyses on the same set of data should be made the norm to assess the robustness of one's conclusions. Our study illustrates this point by showing that there might not be a single appropriate way of analysing a dataset and that multiple, equally plausible, statistical analyses can lead to different conclusions. We also suggest that advanced methods might provide complementary information to infant fNIRS data analysis, but in this study, we have not been able to find enough evidence to suggest relinquishing of more basic methods.

## Acknowledgements

We would like to thank Jun Yin for sharing the scripts for the baseline-corrected averaging approach and Antonia Hamilton for sharing the original scripts for the GLM analyses. We also would also like to thank Oracle for Research staff in Europe, Richard Pitts, and Michael Riley, for their support via Oracle Cloud credits and technical assistance relating to this research.